\documentclass[sn-mathphys]{arxiv}

\usepackage{amsmath}
\usepackage{mathrsfs}
\usepackage[T1]{fontenc} 
\usepackage{animate}
\usepackage{newfloat,caption}

\DeclareFloatingEnvironment[fileext=lop,name=MOVIE]{movie}

\jyear{2022}%

\def\ii{{\rm i}}  \def\ee{{\rm e}}

                                            \def\Rb{{\bf R}}  \def\rb{{\bf r}}       
  
  \def\Lp({L_{\rm p}}       
        
\begin{document}

\title[CDEM: nanoscale imaging of femtosecond plasma dynamics]{Charge dynamics electron microscopy: nanoscale imaging of femtosecond plasma dynamics}

\author[1]{\fnm{Ivan} \sur{Madan}}
\equalcont{These authors contributed equally to this work.}

\author[2]{\fnm{Eduardo~J.~C.} \sur{Dias}}
\equalcont{These authors contributed equally to this work.}

\author[1]{\fnm{Simone} \sur{Gargiulo}}
\equalcont{These authors contributed equally to this work.}

\author[1,3]{\fnm{Francesco} \sur{Barantani}}
\equalcont{These authors contributed equally to this work.}

\author[4]{\fnm{Michael} \sur{Yannai}}

\author[1]{\fnm{Gabriele} \sur{Berruto}}

\author[1]{\fnm{Thomas} \sur{LaGrange}}

\author[1]{\fnm{Luca} \sur{Piazza}}

\author[5]{\fnm{Tom~T.~A.} \sur{Lummen}}

\author[4]{\fnm{Raphael} \sur{Dahan}}

\author[4]{\fnm{Ido} \sur{Kaminer}}

\author[6]{\fnm{Giovanni Maria} \sur{Vanacore}}

\author[2,7]{\fnm{F.~Javier} \sur{Garc\'{\i}a~de~Abajo}}

\author*[1]{\fnm{Fabrizio} \sur{Carbone}}\email{fabrizio.carbone@epfl.ch}

\affil[1]{\orgdiv{Institute of Physics}, \orgname{ École Polytechnique Fédérale de Lausanne}, \orgaddress{ \city{Lausanne}, \postcode{1015}, \country{Switzerland}}}

\affil[2]{\orgdiv{ICFO-Institut de Ciencies Fotoniques}, \orgname{ The Barcelona Institute of Science and Technology}, \orgaddress{\city{Castelldefels (Barcelona)}, \postcode{08860}, \country{Spain}}}

\affil[3]{\orgdiv{Department of Quantum Matter Physics}, \orgname{University of Geneva}, \orgaddress{\street{24 Quai Ernest-Ansermet}, \city{Geneva}, \postcode{1211}, \country{Switzerland}}}

\affil[4]{\orgdiv{Department of Electrical and Computer Engineering}, \orgname{Technion}, \orgaddress{\city{Haifa}, \postcode{32000}, \country{Israel}}}

\affil[5]{\orgdiv{BSSE Single Cell Facility}, \orgname{ETH Zurich}, \orgaddress{\street{26 Mattenstrasse}, \city{Basel}, \postcode{4058}, \country{Switzerland}}}

\affil[6]{\orgdiv{Department of Materials Science}, \orgname{ University of Milano-Bicocca}, \orgaddress{\street{Via Cozzi, 55}, \city{Milano}, \postcode{20126}, \country{Italy}}}

\affil[7]{\orgdiv{ICREA}, \orgname{Instituci\'o Catalana de Recerca i Estudis Avan\c{c}ats}, \orgaddress{\street{ Passeig Llu\'{\i}s Companys, 23}, \city{Barcelona}, \postcode{08010}, \country{Spain}}}
\affil[]{}
\affil[]{}


\abstract{
Understanding and actively controlling the spatio-temporal dynamics of non-equilibrium electron clouds is fundamental for the design of light and electron sources, novel high-power electronic devices, and plasma-based applications. However, electron clouds evolve in a complex collective fashion on nanometer and femtosecond scales, producing electromagnetic screening that renders them inaccessible to existing optical probes. Here, we solve the long-standing challenge of characterizing the evolution of electron clouds generated upon irradiation of metallic structures using an ultrafast transmission electron microscope to record the charged plasma dynamics. Our approach to charge dynamics electron microscopy (CDEM) is based on the simultaneous detection of electron-beam acceleration and broadening with nanometer/femtosecond resolution. By combining experimental results with comprehensive microscopic theory, we provide deep understanding of this highly out-of-equilibrium regime, including previously inaccessible intricate microscopic mechanisms of electron emission, screening by the metal, and collective cloud dynamics. Beyond the present specific demonstration, the here introduced CDEM technique grants us access to a wide range of non-equilibrium electrodynamic phenomena involving the ultrafast evolution of bound and free charges on the nanoscale.
}

\maketitle




\section*{Introduction}\label{sec1}

Modern ultrafast spectroscopy and microscopy strive to explore how electronic and crystal structures evolve on timescales of a few femtoseconds~\cite{Vanacore2016, Sobota2021, Maiuri2020}. However, the complex spatio-temporal dynamics of charge carriers photo-excited/emitted from surfaces
has so far remained largely inaccessible because of the intrinsic difficulty of simultaneously addressing the nanometer and femtosecond scales on which the associated charge and transient near-field dynamics takes place. Understanding such dynamics is essential for the exploration of new physics and the development of applications in high-brightness electron sources in wake-field accelerators~\cite{He2013} and RF/THz-driven emitters~\cite{Musumeci2010,Lange2020,RonnyHuang2016}, ultrafast power electronics~\cite{Samizadeh2020}, plasma X-rays sources~\cite{Miaja-Avila2016, Fullagar2007, Bargheer2004, Sokolowski-Tinten2001,Higashiguchi2006}, plasma tailoring for photon down-conversion~\cite{Nie2018}, and nuclear reactions in laser-generated plasma environments \cite{Wu2018,Gunst2018}. In these contexts, the evolution of plasma is commonly monitored through far-field radiation, and some of its properties are inferred by comparison to numerical simulations \cite{Musumeci2010,Samsonova2019}, with no direct access into microscopic charge or field dynamics on their natural ultrafast nanoscopic scale.

\begin{figure}[h!]
\centering
\includegraphics[width = \linewidth]{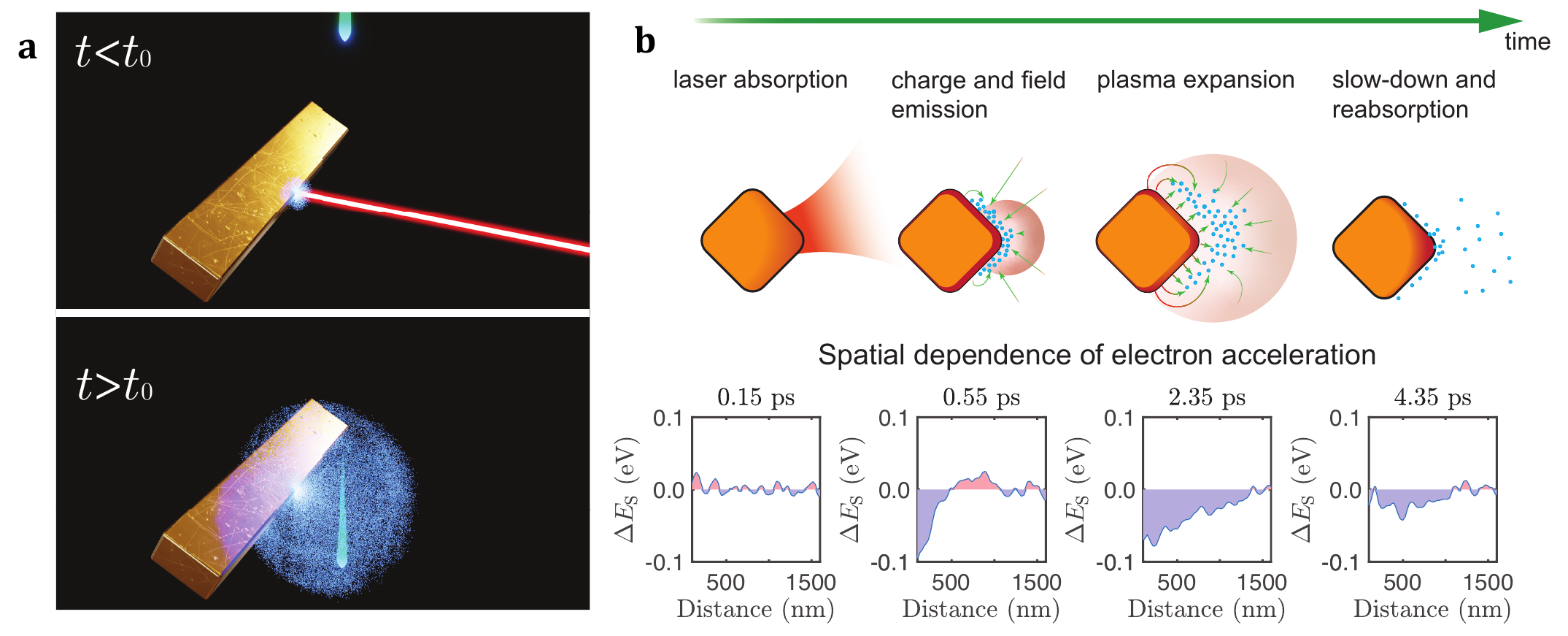}
\caption{\textbf{The CDEM technique and its application to image ultrafast nanoscale plasma dynamics.}
\textbf{a}, Schematics of the studied phenomenon. A laser pulse (50~fs, 800~nm) generates a cloud of photoemitted electrons that is probed by an e-beam pulse ($200$~keV, 600~fs) with a tunable delay time relative to the laser pulse.
\textbf{b}, Differentiated stages (left to right) in the dynamics of the generated electron cloud (upper schemes) and its impact on the transmitted electrons (lower plots): initial laser irradiation; photoemission and THz field generation; explosive phase of rapid charge expansion; and charge density depletion via surface reabsorption. Lower plots show the average e-beam energy change $\Delta E_{\mathrm{S}}$ as a function of e-beam-surface distance at selected delay times (upper labels).
}
\label{fig:microscopies}
\end{figure}

An exemplary scenario involving complex charge dynamics is the process ensuing electron cloud emission from a solid target upon irradiation by high-fluence femtosecond laser pulses (see Fig.~\ref{fig:microscopies}a). The emitted electrons evolve by following distinct stages after light absorption: electron emission, expansion, deceleration, and reabsorption (left to right in Fig.~\ref{fig:microscopies}b). These processes are strongly affected by repulsive Coulomb interactions among electrons and attractive interaction with the screening image charges created on the material surface~\cite{Wendelen2010,Tao2017}. As a result, the charge distribution close to the surface exhibits strong spatial inhomogeneities on the nanometer/femtosecond scales, which remain largely unexplored in experiments \cite{Musumeci2010,Schafer2010} despite their pivotal role in developing potential applications.

Here, we introduce a new approach to access the spatio-temporal dynamics of high-density photo-emitted electron clouds: charge dynamics electron microscopy (CDEM), performed in an ultrafast transmission electron microscope (UTEM) in which $50$-fs, 800-nm laser pulses are used to irradiate a metallic target and $600$-fs electron-beam (e-beam) pulses are probing the dynamics of the emitted electrons (Fig.~\ref{fig:microscopies}a). The spectra of the electron pulses are then recorded as a function of e-beam spatial position and delay time relative to the laser pulses. The emission and subsequent dynamics of the charge cloud generates broadband low-frequency (THz) non-conservative electromagnetic fields, which produce a sizeable overall acceleration of the transmitted e-beam. The dependence of the measured acceleration on e-beam position and delay time relative to the laser pulse reveals a wealth of information on the spatio-temporal dynamics of the electron cloud, as well as its interaction with the emitting material. The entire process involves strong dynamical screening of the exciting laser, ultrafast internal carrier dynamics and thermalization, thermionic and multiphoton photoemission, Coulomb interactions between free-space and image charges, electron-surface recollisions, the generation of low-frequency fields, and the interaction of the latter with the sampling e-beam. We supplement our experiments with a comprehensive microscopic theoretical analysis of these processes in excellent agreement with the measured data, allowing us to conclusively establish four well differentiated stages of charge evolution, as illustrated in Fig.~\ref{fig:microscopies}{b}. 

\section*{Experimental results}\label{sec2}

\begin{figure}[h!]
\centering
\includegraphics[width =\linewidth]{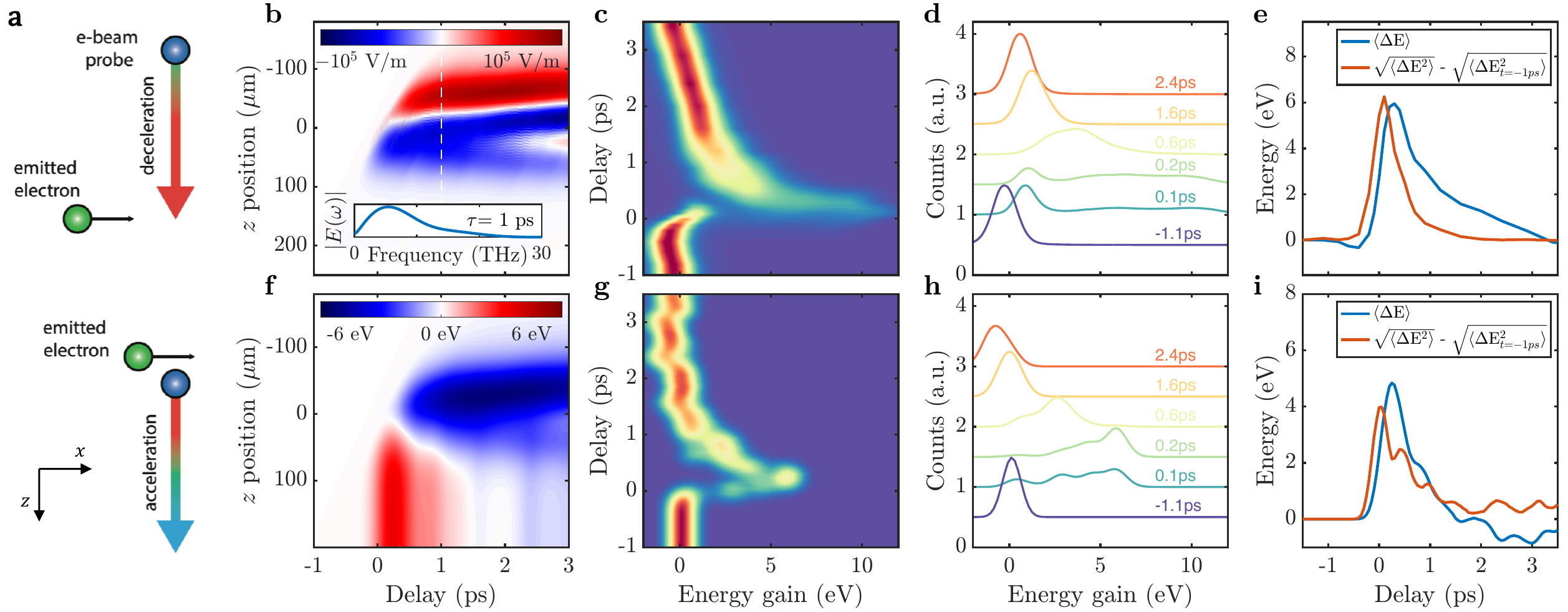}
\caption{\textbf{Ultrafast e-beam interactions in CDEM.}
\textbf{a}, Sketches illustrating e-beam deceleration and acceleration stages, which result in an average net energy change, as well as spectral reshaping.
\textbf{b}, Simulated electric field experienced by the e-beam as a function of delay and position along the trajectory ($z=0$ corresponding to the e-beam leveled with the tip of metal corner in Fig.~\ref{fig:microscopies}) for an e-beam--surface separation of 100~nm. The inset shows the Fourier-transformed electric-field amplitude at a delay of 1~ps (white dashed line), peaking at $5.4$~THz.
\textbf{c}, Transmitted electron spectra as a function of laser-e-beam delay for 100-nm e-beam-surface separation.
\textbf{d}, Profiles extracted from \textbf{c} at selected delays (see labels).
\textbf{e}, Variation of the average e-beam energy and spectrum variance as a function of delay. 
\textbf{f}, Calculated e-beam energy variation as a function of delay and position under the conditions of \textbf{b}.
\textbf{g}, \textbf{h}, \textbf{i}, Numerical simulations based on microscopic theory corresponding to the conditions in \textbf{c}, \textbf{d},  and \textbf{e}, respectively.}
\label{fig:data}
\end{figure}

The main observable in our measurements is the spatial pattern of acceleration experienced by the energetic e-beam probe after passing next to or through the emitted electron cloud. As the latter evolves, it produces time-varying electromagnetic fields that comprise 
low-frequency components interacting with the e-beam (1-10 THz, see inset to Fig.~\ref{fig:data}{b} and Extended Data  Fig.~\ref{fig:SI3}). The acceleration of free electrons by THz fields has been previously investigated using, for example, point-projection electron microscopy \cite{Hergert2021}. However, the CDEM technique performed in an UTEM represents a radical step forward in our ability to probe dense plasmas ($10^{14}~\mathrm{cm}^{-3}$) of different geometries, sizes, and densities with a resolution in the nanometer/femtosecond range over a large field of view.

In our experiment, an electron cloud is photoemitted from a corner of a metal structure, expanding with drift kinetic energies of a fraction of eV.
We find that the acceleration observed in the e-beam is predominantly caused by cloud charge motion along transverse directions, as schematically depicted in Fig.~\ref{fig:data}{a}, while motion parallel to the e-beam contributes negligibly for the cloud velocities observed in our experiment. 
The dynamical character of the interaction is essential. In contrast, for quasi-static charge motion, as explored in deflectometry-based experiments~\cite{Zandi2020,Sun2020,Schafer2010,Scoby2013,Hebeisen2008,Li2010,Raman2009}, the deceleration and acceleration of the probe electron before and after transit are perfectly balanced and produce no net effect. Instead, for rapidly and non-inertially evolving charges, the two contributions are unbalanced and result in a net energy transfer to the e-beam (Fig.~\ref{fig:data}{a}).

A typical temporal evolution of the measured e-beam acceleration is presented in Fig.~\ref{fig:data}c,d, which shows the measured change in the electron spectrum as a function of the delay time relative to the laser pulse for a fluence of 126~$\mathrm{mJ/cm^2}$. The temporal dynamics consists of a strong electron acceleration and spectral broadening at short delays, followed by a slow reduction of the acceleration and, eventually, even deceleration (Fig.~\ref{fig:data}c,d). Qualitatively, as inferred from the schematics in Fig.~\ref{fig:data}{a}, a net acceleration is observed as long as there is a current flowing towards the e-beam. In an intuitive picture, the observed acceleration is the result of the work done on the electron by the electric field ${\bf \mathcal{E}}[{\bf r}_{\rm e}(t),t]$ generated by the cloud electrons and image charges acting along the probe trajectory ${\bf r}_{\rm e}(t)$. The electron energy change is given by the time integral
\begin{align}
\label{DeltaE}
\Delta E= -e {\bf v}_{\rm e}\cdot \int_{-\infty}^{\infty} dt\, {\bf \mathcal{E}} [{\bf r}_{\rm e}(t),t],
\end{align}
where ${\bf v}_{\rm e}$ is the e-beam velocity vector, taken to be approximately constant in the evaluation of Eq.~(\ref{DeltaE}). This expression, which represents the work done on a classical point-particle electron, can be rigorously derived from a quantum-mechanical treatment of the e-beam when the external THz field varies negligibly during the interaction time defined by $\tau_{\rm interaction}\sim L/v_{\rm e}$, which is $\sim1$\,ps for an effective interaction length $L\sim200\,\mu$m (see Fig.~\ref{fig:data}b) along the beam direction (see \nameref{secMethod}).

To better understand the origin of the acceleration and estimate the effect of free and image charges, experimental geometry, and material properties, we compare the measured data with simulations based on a comprehensive account of the different microscopic physical processes involved in the generation and evolution of the electron cloud, as well as its interaction with the probing electron (see details of the theory in \nameref{secMethod} and S.M.~\cite{suppmater}). Figure~\ref{fig:data}{b} shows the calculated electric-field component parallel to the e-beam  as a function of both (i) the delay between laser and e-beam pulses and (ii) the position along the electron trajectory, while the inset shows its spectral decomposition at 1 ps delay. The resulting delay- and position-dependent variation of the e-beam energy is shown in Fig.~\ref{fig:data}{f}, as obtained from Eq.~(\ref{DeltaE}) by setting the upper integration limit to a finite time corresponding to each electron-probe position (see also Extended Data Fig.~\ref{fig:SI3}). We observe that the electric field rapidly decays away from the metal and becomes negligible at distances $>100\,\mu$m, beyond which the e-beam energy remains unchanged (i.e., at the value recorded in experiment). Accounting for the finite e-beam pulse duration, we also calculate the e-beam spectrum as a function of probe delay (Fig.~\ref{fig:data}g,h), finding remarkable qualitative and quantitative agreement with experiment.

To quantitatively capture both the amplitude of the acceleration and the observed timescales, two different emission processes need to be considered: thermionic, due to the raise in electron temperature, which stays elevated during a ps time scale; and three-photon photoemission, which occurs within the $50$~fs duration of the laser pulse (see \nameref{secMethod} for details). At each time step in the simulation, the force exerted on every individual electron by the remaining electrons and their associated induced surface image charges is evaluated, and its position and velocity are evolved accordingly (see Extended Data Fig.~\ref{fig:SI2} for details on the plasma charge dynamics). Partial electron absorption upon recollision with the metal surface is also accounted for. The net energy variation of a probing electron after traversing the plasma is then calculated from Eq.~(\ref{DeltaE}), with the electric field obtained by summing the contributions from all emitted cloud electrons and their associated image charges, including the effect of surface geometry and retardation, and further averaging over the electron wave-packet density profile (see S.M.~\cite{suppmater}). Our simulations reveal that the contribution of the interactions with the electron cloud and the image charges produce two components of similar amplitude but with opposite signs (see Extended Data Fig.~\ref{fig:SI1}c). However, image charges are constrained to the material surface, so that their contribution is weaker than that of free-space charges. Effectively, the e-beam probe sees an effective dipolar field, with the dipole oriented nearly transversely with respect to the beam direction. 

As shown in Fig.~\ref{fig:data}{e}, our CDEM measurements unveil two main time scales: (i) fast plasma creation by thermionic and photoemission processes
occurring faster than the electron pulse duration (FWHM $\simeq600$~fs); and (ii) plasma dynamics driven by space and image charges, which manifests as a gradual relaxation of the electron energy shift $\Delta E$ over 1-2~ps. During the former, in addition to the net acceleration, we observe a substantial broadening of the electron spectrum, which we quantify by computing the second moment $\sqrt{\langle\Delta E^2\rangle}$. This experimental broadening, which is also captured in our simulations, exhibits a maximum at the delay for which we encounter the largest variation of the average acceleration (Fig.~\ref{fig:data}e,i), so that peak broadening and peak acceleration are mutually delayed by $\sim 200$~fs.

\begin{figure}[h!]
\centering
\includegraphics[width = 1\textwidth]{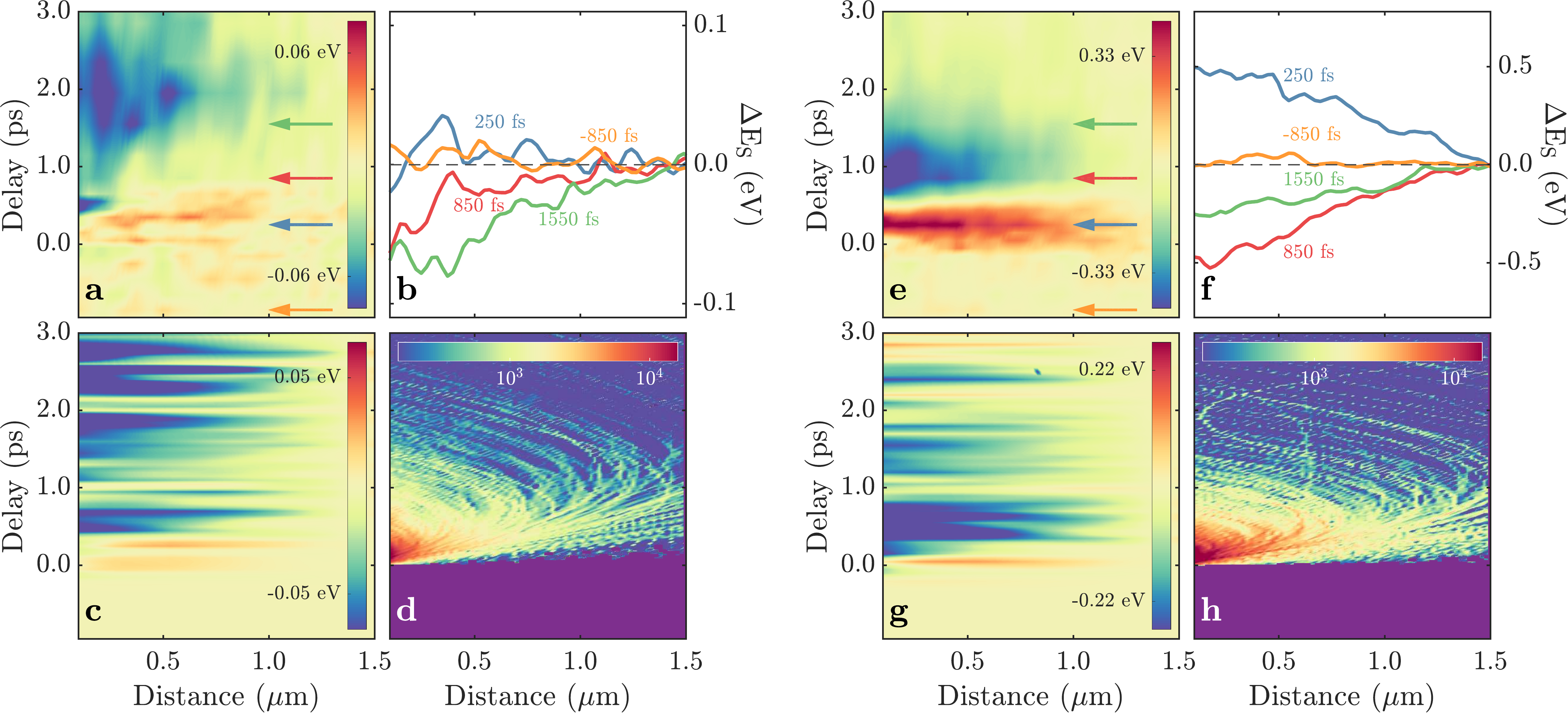}
\caption{\textbf{Spatio-temporal evolution of the average electron acceleration.}
\textbf{a}, Measured space- and time-resolved variation of electron acceleration $\Delta E_{\mathrm{S}}$ under excitation with 126~$\mathrm{mJ/cm^{2}}$ laser pulse fluence.
\textbf{b}, Selected spatial profiles of $\Delta E_{\mathrm{S}}$ from \textbf{a}.
\textbf{c}, Numerical simulation under the conditions of \textbf{a}.
\textbf{d}, Spatio-temporal evolution of the electron density extracted from the simulation in \textbf{c}.
\textbf{e}, \textbf{f}, \textbf{g}, \textbf{h}, Same as \textbf{a}, \textbf{b}, \textbf{c}, \textbf{d}, respectively, but for 189~$\mathrm{mJ/cm^{2}}$ laser pulse fluence.
}
\label{fig:analysis}
\end{figure}


\section*{Spatially-resolved charge cloud evolution} \label{sec3}
The presence of free-space and image charges drastically affects the expansion and evolution of the electron cloud ~\cite{Riffe1993, Wendelen2010, Wendelen2012}. For example, charge expansion is close to ballistic at low fluences, when emitted charge densities are small. In contrast, when the cloud reaches large densities, newly emitted electrons are trapped closer to the surface due to the strong Coulomb repulsion by previously emitted electrons~\cite{Wendelen2010,Tao2017}, causing the number of electrons that permanently escape the photo-excitation region to be drastically reduced down to only a fraction $\sim10^{-6}-10^{-8}$ of the total emission~\cite{Wendelen2010}. 
Those that acquire sufficient velocity to escape the photo-excitation region can be investigated by electron detectors and imaged with electron-deflection-based techniques ~\cite{Zandi2020,Sun2020,Schafer2010,Scoby2013,Hebeisen2008,Li2010,Raman2009}, while in this work we provide insight into the previously inaccessible high-density electron cloud that is eventually reabsorbed during the first few picoseconds after emission.

The spatial extension of the expanding charged plasma, its initial velocity, and the deceleration due to interaction with the image charges are all pivotal elements of information that can be extracted by studying the spatial variation of the e-beam probe acceleration in CDEM. 
In Fig.~\ref{fig:analysis}, we present the spatial variation of the e-beam energy change as a function of delay time and beam position:
$\Delta E_\mathrm{S}(t,d) = \langle E \rangle (t,d) - \langle E \rangle (t,d_{\mathrm{max}})$, referred to the e-beam-target separation for the maximum explored distance $d_{\mathrm{max}}=1.5~\mu$m. This allows us to precisely follow the spatial dynamics developing over the average acceleration. Experimental results for 126 and 189~$\mathrm{mJ/cm^{2}}$ laser pulse fluence are shown in Fig.~\ref{fig:analysis}a,e, respectively. Close to the metal surface, $\Delta E_{\mathrm{S}}$ is positive at early delay times, while it becomes negative at later delay times. This negative feature is characterised by faster rise and decay times when irradiating with a larger fluence (see selected profiles in Fig.~\ref{fig:analysis}b,f).

Figure~\ref{fig:analysis}c,g shows the corresponding numerical simulations for $\Delta E_\mathrm{S}$, qualitatively reproducing the experimental features, including their timescales and variation with fluence. 
A comparison with the evolution of the emitted electron density (Fig.~\ref{fig:analysis}d,h) allows us to gain further insight into the relation between the observed behavior of $\Delta E_{\mathrm{S}}$ and the charge dynamics. At short delay times, the emitted charge cloud is localized close to the surface, and the spatial variation of the observed acceleration is reminiscent of the power-law dependence of the near-field THz component. At the same time, the expanding charge density causes a decrease in the electron beam acceleration due to partial screening, as manifested in the negative region in Fig.~\ref{fig:analysis}a,e, whose onset permits us to experimentally determine the initial charge expansion velocity as $\sim$ 1.2 nm/fs.

Due to the interaction with image charges, most of the emitted electrons slow down in the immediate vicinity of the surface and are eventually reabsorbed. This is confirmed upon inspection of individual particle trajectories in our theory (Fig.~\ref{fig:analysis}d,h), which bend to the surface and eventually collide with it within a few hundred femtoseconds, while those that acquire higher speed in the initial stage are able to escape the surface-neighboring region.
Electrons that are colliding with the surface do not observably influence the e-beam probe spectrum because of their reduced speed and the cancelling fields originating in proximal positive (images) and negative (electron) charges.

Expansion and reabsorption of the electron cloud result in a reduction of the cloud density (see Fig.~SI2a), which leads in turn to a gradual depletion of the negative $\Delta E_{\mathrm{S}}$ region close to the sample surface on a 1-2 ps timescale (see Fig.~\ref{fig:analysis}a,c,e,g).

\section*{Perspective}\label{sec4}

\begin{figure}[h!]
\centering
\includegraphics[width = 1\textwidth]{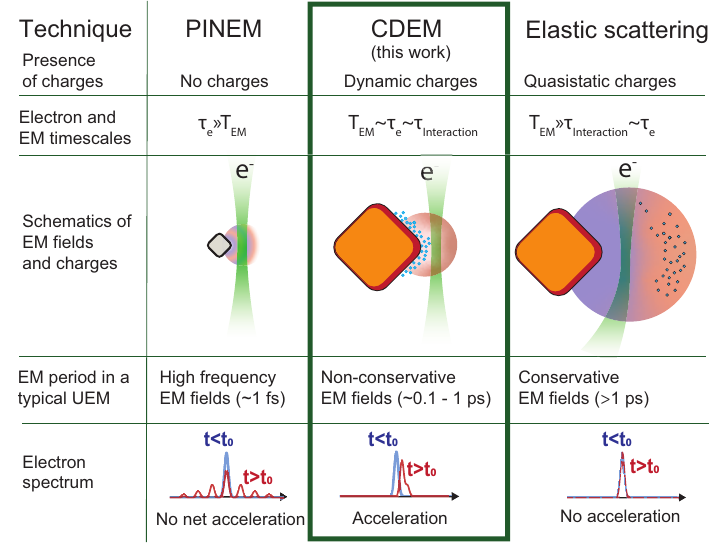}
\caption{\textbf{Comparison between different ultrafast electron microscopy techniques and the corresponding interactions between the probe electrons and electromagnetic near fields.} The table compares relevant parameters and the main differences between the techniques, relating timescales and experimental observables. EM: electromagnetic; UEM: ultrafast electron microscope; CDEM: charge dynamics electron microscopy; PINEM: photon-induced near-field electron microscopy.
}
\label{fig:table}
\end{figure}

In perspective, CDEM covers a previously unexplored regime of ultrafast interaction between e-beams and near fields, as emphasized in Fig.~\ref{fig:table}, which compares CDEM both to photon-induced near-field electron microscopy (PINEM)~\cite{Barwick2009, GarciadeAbajo2010, Park2010} and to electron microscopy methods based on elastic electron-field interactions \cite{Zandi2020,Sun2020,Schafer2010,Scoby2013,Hebeisen2008,Li2010,Raman2009}. The latter (Fig.~\ref{fig:table}, right column) involves an optical cycle of the electromagnetic field $T_{\rm EM}$ that is large compared with both $\tau_{\rm interaction}$ and the electron pulse duration $\tau_{\rm e}$. This regime includes Lorentz transmission electron microscopy, electron holography, deflectometry, and shadowgraphy, which are sensitive to slow quasi-static conservative electric fields ~\cite{Sun2020, Centurion2008}. On the opposite extreme, PINEM (Fig.~\ref{fig:table}, left column) capitalizes on the effect of rapidly oscillating optical fields ($T_{\rm EM}\ll\tau_{\rm e}$), which show up as inelastic peaks in the electron spectrum at multiples of the photon energy, usually configuring a symmetric spectrum (for nearly monochromatic illumination) with respect to the elastic peak due to the stimulated nature of the process and the large occupation number of the involved laser-driven optical modes. Under exposure to monochromatic fields, the net e-beam energy change in PINEM is zero, just like in elastic diffraction techniques. This is one key aspect by which CDEM deviates from other techniques: the electron spectrum is asymmetric, producing a sizeable e-beam energy change. Indeed, the intermediate regime in which the interaction, electron-pulse, and optical-cycle times are commensurate (Fig.~\ref{fig:table}, central column) is where CDEM belongs ---a natural domain to extract spatio-temporal information on the probed fields and associated sources. A unified, rigorous quantum-mechanical formalism can simultaneously capture all three regimes with a relatively simple theory (see S.M.~\cite{suppmater} for a detailed derivation), under the approximation that the kinetic energy of the incident probe electron largely exceeds the energy change due to the interaction, as is the case here. In such scenario, the incident wave function is multiplied by a factor involving the exponential of an action (the integrated field along the probe trajectory), which becomes an energy comb for monochromatic fields (i.e., the PINEM limit); in contrast, the same factor reduces to the energy shift given by Eq.~(\ref{DeltaE}) in the classical limit \cite{suppmater}.

In contrast to PINEM and elastic scattering, the CDEM approach allows us to follow the formation and evolution of dense plasma with femtosecond/nanometer space/time resolution. From a technological viewpoint, access to spatially-resolved information offers the possibility to develop customized nanostructures that can be optimized to operate on ultrafast timescales \cite{Samizadeh2020}. Also, from a material science perspective, CDEM enables the investigation of image charge dynamics and screening timescales in out-of-equilibrium nanostructured materials, allowing us to map spatial inhomogeneities such as the formation of domains following a phase transition.

\section*{Discussion and outlook}

Through the insight gathered from CDEM on free-space electron clouds, combined with a predictive degree of theoretical insight, we introduce a powerful tool for the quantitative optimization of electron sources operating under extreme space-charge conditions. This has potential application in nanopatterned radiofrequency-gun electron emitters, where 100~nm-sized features have been demonstrated to produce 100-fold electron yield enhancement~\cite{Li2013}. Similarly, periodic arrays of electron-plasma emitters can drastically improve the emission efficiency on the femtosecond scale by operating in a high-plasma-density regime exceeding critical values by orders of magnitude~\cite{Samsonova2019}. CDEM is an ideal tool to diagnose such super-critical plasma, providing nanometer/femtosecond space/time-resolved imaging to optimize geometrical and compositional parameters. Similar benefits are expected in the development of plasma-based high-efficiency X-ray sources, nanoelectronic devices, and nuclear or astrophysics-in-a-lab experiments \cite{Zhang2021}. 

The intense nanoscale THz fields produced by the diagnosed plasma hold strong potential for use in the spatial, angular, and spectral compression of e-beams, enabling finer spatio-temporal control with respect to traditional THz-based approaches \cite{Ehberger2019,Kealhofer2016}. CDEM could thus be applied to manipulate the wave function of free-electrons in ways that existing techniques such as PINEM cannot. In addition, the photon statistics of the THz field associated with the out-of-equilibrium plasma remains as a fundamental question \cite{Dahan2021} that cannot be addressed with conventional quantum-optics techniques because of the limited speed and sensitivity of available THz photodetectors \cite{Kitaeva2014,Kutas2020,Prudkovskii2021}. CDEM is thus offering a viable approach to characterize the statistics of near-field photons at low frequencies.

\section*{Methods}\label{secMethod} 
\subsection*{Sample preparation and UTEM experiments}

For the experiments reported, we used a copper 100 Mesh PELCO grid. The grid was tilted by $45^\circ$ with respect to the $z$ direction (parallel to the TEM column axis) in order to expose the corner of a rectangular copper rod with a cross section of $\sim50\times25\,\mu$m$^2$ (see Extended Data Fig.~\ref{fig:SI1}). The edge of the rod corner exhibited a radius of curvature of 4~$\mu$m, as estimated from SEM micrographs. The sample was positioned such that only one of the edges of the rectangular rod was illuminated by the laser pulse.

To generate the charged plasma, we irradiated the copper rod with near-infrared laser pulses of 1.55~eV central photon energy and 50~fs temporal duration at a repetition rate of 100~kHz. Light entered the microscope through the zero-angle port and was focused under normal incidence on the copper rod via an external plano-convex lens. In such a geometry, the light beam was also perpendicular with respect to the electron propagation direction.

The dynamics of the photoemitted electrons was then probed by means of electron pulses with a temporal duration of about 600~fs, with controlled delay between electron and laser pulses. All the experiments were performed in a modified JEOL 2100 TEM microscope at an acceleration voltage of 200~kV~\cite{Piazza2013, Piazza2012}. The probe electrons were generated by illuminating  a LaB$_6$ cathode with third-harmonics UV light at 4.65~eV photon energy. 

Our transmission electron microscope was equipped with EELS capabilities coupled to real-space imaging. Energy-resolved spectra were recorded using a Gatan-Imaging-Filter (GIF) camera operated with a 0.05~eV-per-channel dispersion setting and typical exposure times of the CCD sensor from 30 to 60 s. 
For the acquisition of space-energy maps (see Fig.~\ref{fig:analysis}), special care was devoted to sample alignment. The copper rod was adjusted to be parallel to the energy dispersion direction and placed at the edge of the spectrometer entrance aperture.

The acquired position-dependent spectra were analysed as a function of delay between the laser and electron pulses, with the time-zero being determined as the peak of PINEM signal observed within 100~nm close to the sample surface at relatively low fluence ($\sim$50 mJ/cm$^2$). Camera noise and signal from cosmic events was reduced by applying a median filtering. Distortions of the spectrometer were corrected by aligning the spectrum according to the negative delay energy-space spectrographs (-2~ps). First and second moments of the spectrum were calculated in a reduced energy window, which is taken 10~eV larger than the region in which the electron signal is above 10\% of the peak value (i.e., the maximum value among all delays and positions measured for a given fluence). This procedure helped to reduce contributions from the CCD background noise.

\subsection*{Classical limit for the energy loss experienced by a free electron traversing an optical field}

We derive a classical limit for the interaction between a collimated free electron and a classical electromagnetic field starting from a quantum-mechanical expression that bears general validity in the nonrecoil approximation.

Under the experimental conditions, the free electron probe has a small energy spread relative to its average kinetic energy both before and after interaction. We can therefore adopt the nonrecoil approximation \cite{GarciaDeAbajo2021} and introduce the interaction with a classical field through the Hamiltonian $(ev/c)A_\mathrm{z}$, where $v$ is the electron velocity and $A_\mathrm{z}$ is the vector potential component along the beam direction $z$ in the Coulomb gauge, for which the scalar potential vanishes within the vacuum space traversed by the electron. We further consider a finite interaction region, in which $v$ is assumed to remain constant, such that the wave function depends on the longitudinal coordinate $z$ and time $t$ only through $z-vt$. Under these conditions, starting from an incident electron wave function $\psi^0(z,t)$, the post-interaction wave function reduces to \cite{GarciaDeAbajo2021, Vanacore2018}
\begin{align}
\psi(z,t)=\psi^0(z,t)\,\exp\bigg\{-\frac{\ii e v}{\hbar c}\int_{-\infty}^\infty dt'\, A_\mathrm{z}(z-vt+vt',t')\bigg\}, \label{psi1}
\end{align}
where an implicit dependence on transverse coordinates $(x,y)$ is understood.

The evaluation of Eq.~\eqref{psi1} for monochromatic fields reduces the exponential factor to a well-known sum over energy sidebands that accurately describes experimentally observed PINEM spectra \cite{Vanacore2018}. In contrast, in the present work the electron is exposed to external fields comprising components whose optical cycles are long compared to the interaction time $L/v_{\rm e}$ (see main text). It is then pertinent to Taylor-expand the slowly varying vector potential $A_\mathrm{z}(z-vt+z',t')$ around small values of $z-vt$, assuming the centroid of the electron wave packet to follow the trajectory $z=vt$. The independent term in this expansion contributes with an overall phase $\varphi=-(ev/\hbar c)\int_{-\infty}^\infty dt\, A_\mathrm{z}(vt,t)$ that does not affect the transmitted electron spectrum. Retaining only the linear term in $z-vt$, Eq.~\eqref{psi1} reduces to
\begin{align}
\psi(z,t)=\psi^0(z,t)\,\ee^{\ii\varphi}\exp\big\{\ii(\Delta E/\hbar)\,(z/v-t)\big\}, \label{psi2}
\end{align}
where
\begin{align}
\Delta E
&=-\frac{ev}{c}\int_{-\infty}^\infty dz'\, \partial_{z''}A_\mathrm{z}(z'',z'/v)\big\vert_{z''=z'} \nonumber\\
&=-\frac{e}{c}\int_{-\infty}^\infty dz'\, \big[v\partial_{z'}A_\mathrm{z}(z',z'/v)-\partial_tA_\mathrm{z}(z',t)\big\vert_{t=z'/v}\big] \nonumber\\
&=-e\int_{-\infty}^\infty dz'\,\mathcal{E}_\mathrm{z}(z',z'/v), \label{DE}
\end{align}
represents the energy change experienced by a classical point electron moving along the noted trajectory. In the derivation of this expression, the first term in the second line cancels upon integration by parts for a field of finite extension along the electron trajectory (i.e., localized at the interaction region), and we have identified $-(1/c)\partial_tA_\mathrm{z}(z,t)=\mathcal{E}_\mathrm{z}(z,t)$ with the electric field component along the beam direction to obtain the third line. In summary, the wave function in Eq.~\eqref{psi2} is the incident one multiplied by an irrelevant phase factor $\ee^{\ii\varphi}$ as well as by a plane wave $\ee^{\ii(\Delta E/\hbar)\,(z/v-t)}$ representing a rigid shift in energy by $\Delta E$ (and a corresponding change in momentum by $\Delta E/v$ within the nonrecoil approximation). From Eq.~\eqref{DE}, we then recover Eq.~\eqref{DeltaE} by setting $z'=vt$. Corrections of higher-order terms in the aforementioned Taylor expansion may become relevant for electron wave-packet durations similar or larger than either the optical cycle or the temporal extension of the external field.

\subsection*{Numerical simulations}

In this section, we describe the main aspects of the theoretical model employed to simulate the experimental results presented in this work. Additional details can be found in the SI.

First, we model the temperature dynamics $T(t,s)$ in the inner surface of the copper bar as a function of time $t$ and surface position $s$ using the two-temperature model (see Extended Data Fig.~\ref{fig:SI1}). The pump illumination is introduced through the near-field distribution calculated in the inner surface through the boundary-element method.

We then model electron emission as a function of local temperature $T$ from two different channels: thermionic emission, due to the heightened temperature of the surface electrons, which extends over a few ps and we evaluate using a surface-barrier model; and three-photon photoemission, resulting from the absorption of three photons by one electron during the duration of the pumping $<100$ fs, calculated using the Fowler-Dubridge model (see SI for details).

Combining these results, we simulate the emission of photoelectrons at each instant of time and surface position, which gives rise to a density of photoemitted electrons $\rho_e(\Rb,t)$ as a function of spatial position $\Rb$ and time $t$. The evolution of the plasma density is then simulated by discretizing time and considering, at each time step, the force acting on each of the photoemitted electrons by all the remaining ones, as well as the interaction with the copper bar. The latter is introduced by rigorously accounting for the accumulation of positive image charges at the copper surface due to the presence of the negatively-charged electrons in its vicinity. The position and velocity of each electron is then evolved according to the net force exerted on it. The eventual collision of the photoemitted electrons with the surface gives rise to partial reabsorption according to the barrier model, as well as specular reflection of the non-absorbed electrons. This procedure allows us to determine $\rho_e(\Rb,t)$ for the full duration of the simulation (see Extended Data Fig.~\ref{fig:SI2}).

Finally, we calculate the energy variation by a probe electron passing with velocity $\mathbf{v}_{\rm e}$ at a distance $b$ from the copper bar along a trajectory $\rb_{\rm e}(t)=\mathbf{r}_0+\mathbf{v}_{\rm e}(t+\tau)$, where $\rb_0$ is the nearest point to the copper bar and $\tau$ is the delay of the probe electron with respect to the laser pump. At each time $t$, we calculate the electric field ${\bf \mathcal{E}}$ at the probe electron position $\rb_{\rm e}(t)$ generated by all of the plasma electrons and their corresponding surface charges, taking into account retardation effects and averaging over the time-duration of the electron wave packet (see SI for additional details). The net energy variation by the probe electron, which is a function of $b$ and $\tau$, is finally obtained by using Eq.~\eqref{DeltaE}.

\backmatter

\bmhead{Acknowledgments}
We thank Ivo Furno and Paolo Ricci for insightful discussions, Veronica Leccese for sample dimensions characterisation and Yassine Benhabib for help with graphics.
This work is supported in part by European Union (Horizon 2020 Research and Innovation Program under grant agreement No 964591 SMART-electron), the European Research Council (ERC Advanced Grant 789104-eNANO and ERC Staring Grant 851780-NanoEP), the Spanish MICINN (PID2020-112625GB-I00 and SEV2015-0522), the Catalan CERCA Program, the Generalitat de Catalunya, and Google Inc. M.Y. and R.D. are partially supported by the VATAT Quantum Science and Technology scholarship.



\bmhead{Competing interests:} The authors declare no competing interests.

\bmhead{Availability of data and materials:} All data are available in the main text or the supplementary materials.

\bmhead{Additional information}
\bmhead{Supplementary information:} The online version contains supplementary material available at \cite{suppmater}.


\bibliography{MyCollection.bib}


\setcounter{figure}{0}
\renewcommand{\figurename}{Extended Data Fig.}
\renewcommand{\thefigure}{\arabic{figure}}
\newpage

\begin{figure}[ht]
\centering
\includegraphics[width=\textwidth]{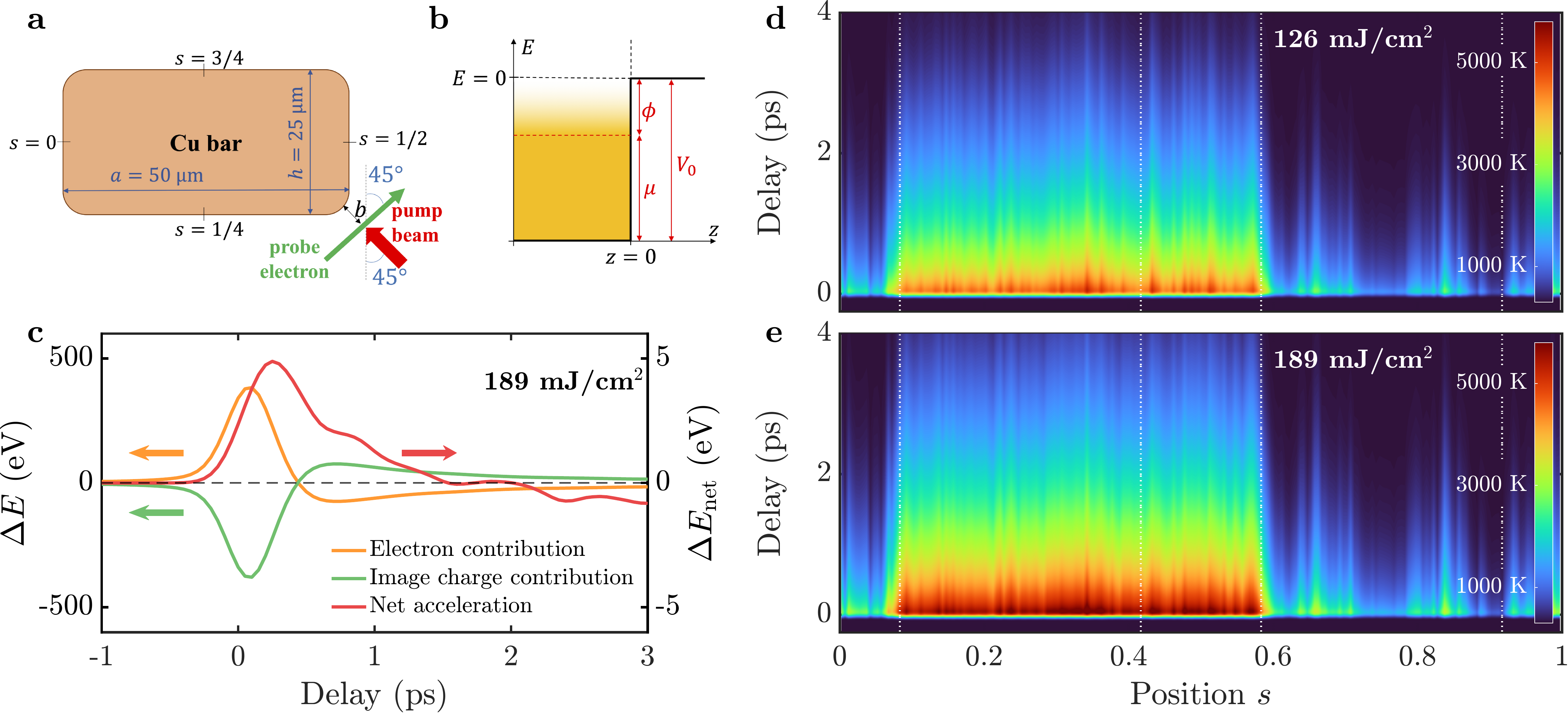}
\caption{\textbf{Numerical simulation details.} \textbf{a}, Scheme of the simulated copper bar cross section, where the characteristic lengths are indicated. The variable $s$ represents the surface parameterization coordinate, ranging from 0 to 1 along its perimeter. The pump beam and electron probe directions are also indicated. Also, $b$ denotes the impact parameter associated to the electron trajectory. \textbf{b}, Potential barrier associated with the conduction band of the metal and magnitudes involved in the description of the electron emission process. Here, $E$ is the electron energy, $V_0$ is the total barrier height, $\mu$ is the chemical potential, $\phi=V_0-\mu$ is the work function, and $z$ is the position-dependent direction perpendicular to the surface. \textbf{c}, Individual contribution to the probe-electron energy change by the image (green) and emitted (orange) charges, and corresponding net energy change (red, right vertical scale), all for an impact parameter $b$=1~$\mu$m. \textbf{d}, \textbf{e}, Spatio-temporal map of the electron temperature at the copper surface upon pump laser excitation, for two fluence values ($126$ and $189$~mJ/cm$^2$). The dashed vertical lines correspond to the parameterized position of the bar corners. Details of the simulations are described in S.M.}
\label{fig:SI1}
\end{figure}

\newpage

\begin{figure}[ht!]
\centering
\includegraphics[width=\textwidth]{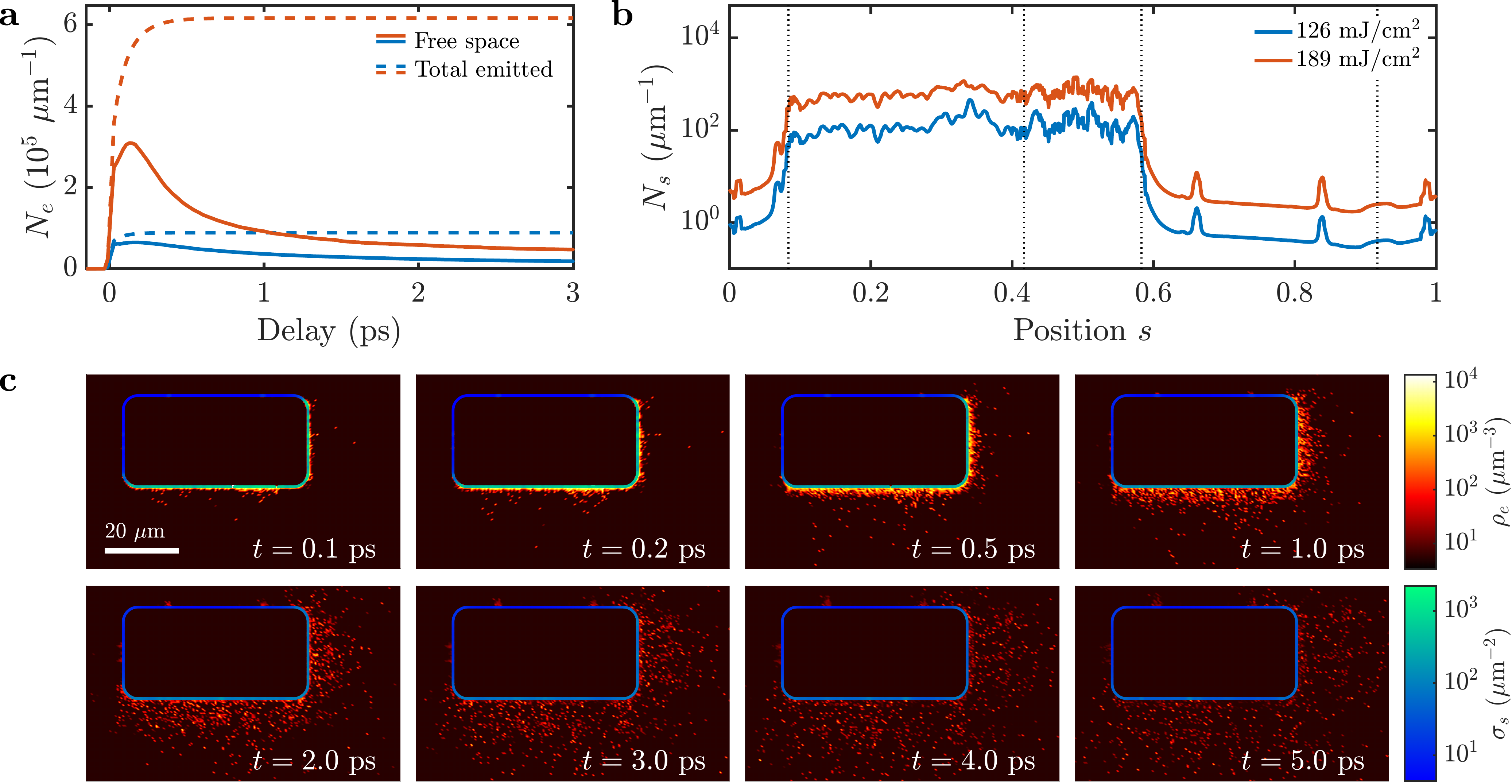}
\caption{\textbf{Plasma charge dynamics.} \textbf{a}, Dynamics of the integrated density of emitted charges in free space (solid curves) and total charges (i.e., accumulated emission, dashed curves) for two fluence values ($126$ and $189$~mJ/cm$^2$) as a function of delay defined with respect to the center of the pump laser pulse.  \textbf{b}, Spatial distribution along the surface of the image charge density (at the time instant when the integrated density is maximum), for the same fluences as in \textbf{a}. The dashed vertical lines correspond to the parameterized position of the bar corners. \textbf{c}, Snapshots at selected delay instants (see labels) of the thermally emitted electron distribution $\rho_e$, and associated surface density distribution $\sigma_s$ along the bar surface.Details of the simulations are described in S.M.}
\label{fig:SI2}
\end{figure}

\newpage

\begin{figure}[ht]
\centering
\includegraphics[width=\textwidth]{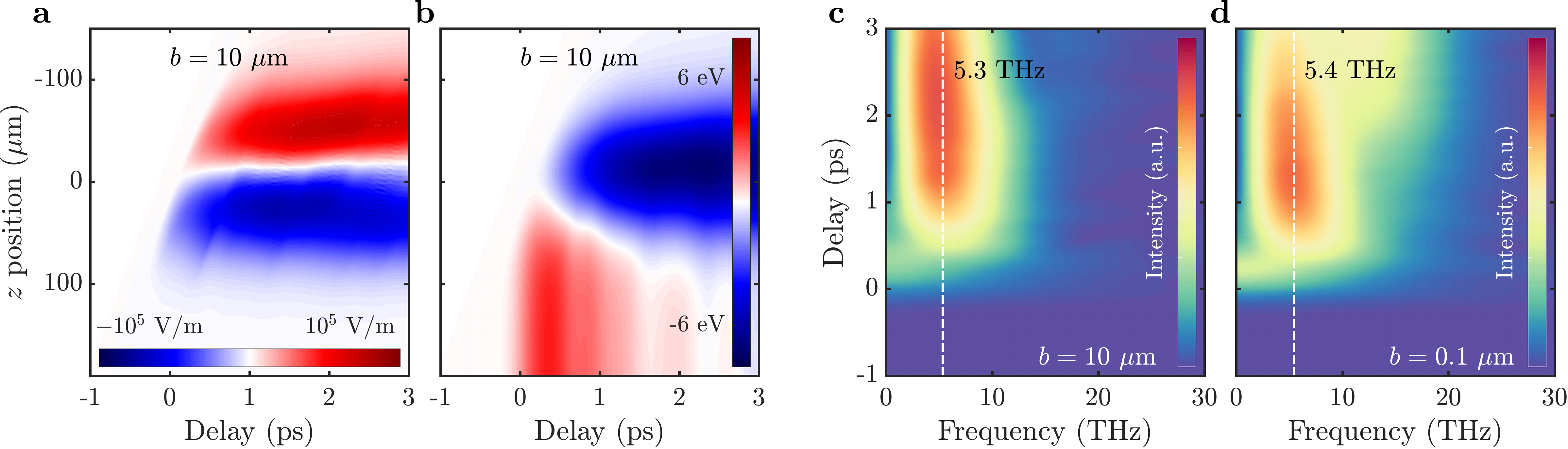}
\caption{\textbf{Energy variation of the probe electron.} \textbf{a}, Electric field at the frame of the probe electron along its trajectory as a function of position along the trajectory (vertical axis, with 0 representing the point closest to the copper bar) and delay between the e-beam and pump-laser pulses. The electric field is chosen to be positive if it points in the same direction as the electron velocity. \textbf{b}, Analogous to panel a, but showing the electron energy variation along its motion. \textbf{c} and \textbf{d}, Fourier transform of the temporal profile of the electric field for impact parameters of $10~\mathrm{\mu m}$ and $0.1~\mathrm{\mu m}$ (see labels). All the results are calculated for an incident fluence of 189 mJ/cm$^2$. \textbf{a} and \textbf{b} are calculated for an impact parameter of $10~\mathrm{\mu m}$.Details of the simulations are described in S.M.}
\label{fig:SI3}
\end{figure}

\end{document}